\documentclass[showpacs,twocolumn,preprintnumbers,amsmath,amssymb,superscriptaddress,nofootinbib]{revtex4}
\usepackage{graphicx}
\usepackage{bm}
\usepackage{ulem}

\begin{document}

\newcommand{\simgt}{\lower.5ex\hbox{$\; \buildrel > \over \sim \;$}}
\newcommand{\simlt}{\lower.5ex\hbox{$\; \buildrel < \over \sim \;$}}
\newcommand{\bmf}[1]{\mbox{\boldmath$#1$}}
\newcommand{\sang}[1]{\left\langle#1\right\rangle}

\title{Measuring the distance-redshift relation with the
  cross-correlation of \\
  gravitational wave standard sirens and galaxies}

\author{Masamune Oguri}
\affiliation{
Research Center for the Early Universe, University of Tokyo, Tokyo 113-0033, Japan
}
\affiliation{
Department of Physics, University of Tokyo, Tokyo 113-0033, Japan 
}
\affiliation{
Kavli Institute for the Physics and Mathematics of the Universe 
(Kavli IPMU, WPI), University of Tokyo, Chiba 277-8582, Japan 
}

\date{\today}

\begin{abstract}
Gravitational waves from inspiraling compact binaries are known to be
an excellent absolute distance indicator, yet it is unclear whether
electromagnetic counterparts of these events are securely identified
for measuring their redshifts, especially in the case of black
hole-black hole mergers such as the one recently observed with the
Advanced LIGO. We propose to use the cross-correlation between spatial
distributions of gravitational wave sources and galaxies with known
redshifts as an alternative means of constraining the
distance-redshift relation from gravitational waves. In our analysis,
we explicitly include the modulation of the distribution of
gravitational wave sources due to weak gravitational lensing. We show
that the cross-correlation analysis in next-generation observations
will be able to tightly constrain the relation between the absolute
distance and the redshift, and therefore constrain the Hubble constant
as well as dark energy parameters. 
\end{abstract}
\maketitle

\section{Introduction}
\label{sec:introduction}

The observation of the gravitational wave (GW) signal GW150914 by
Advanced Laser Interferometer Gravitational-Wave Observatory (LIGO)
opened up the possibility of using GWs as cosmological and
astrophysical probes \cite{Abbott:2016blz,Abbott:2016nhf,TheLIGOScientific:2016htt}.
GW150914 is a GW signal from a pair of merging back holes (BHs) at
$z\sim 0.1$, with a mass of each BH of $M\sim 30M_\odot$ , which were
inferred by fitting the observed waveform with numerical relativity
waveforms. The discovery of such a BH-BH merger event implies that
BH-BH mergers may be much more ubiquitous than previously thought.

GWs from inspiraling compact binaries, which sometimes referred as
standard sirens, are potentially a very powerful cosmological probe,
because they can determine the absolute distances to the GW sources
\cite{Schutz:1986gp}. Assuming general relativity, one can obtain
information on masses of inspiraling and merging objects from the
shape of the waveform, which then determines the absolute strain
amplitude of GWs emitted from these objects. Thus the comparison of
the observed strain amplitude enables a direct measurement of the
luminosity distance to the GW source. Indeed, the luminosity distance
to GW150914 was estimated to be $410^{+160}_{-180}~$Mpc using this
technique \cite{Abbott:2016blz}. With the redshift information from
independent observations of an electromagnetic (EM) counterpart, one
can directly constrain the absolute distance-redshift relation, and
hence obtains constraints on cosmological parameters including the
Hubble constant and dark energy parameters  
\cite{Holz:2005df,Dalal:2006qt,Cutler:2009qv,Nissanke:2009kt,Shapiro:2009sr,Hirata:2010ba,Hilbert:2010am,Nishizawa:2010xx,Taylor:2012db,Camera:2013xfa,Arabsalmani:2013bj,Tamanini:2016zlh}.

However, it is unclear whether EM counterparts can really be
observed. In the case of GW150914, while a hard X-ray emission that
might possibly be associated with GW150914 was detected with Fermi
Gamma-ray Burst Monitor \cite{Connaughton:2016umz}, no confirmed EM
counterpart is reported. Since mergers of BHs are not expected to
have EM counterparts, this association of the hard X-ray emission, if
real, requires special explanations such as a BH-BH merger in a dense
environment  (e.g. \cite{Loeb:2016fzn}). If a bulk of BH-BH mergers do
not have EM counterparts, their use as a cosmological probe will be
limited.  

It has been argued that GWs from compact binary mergers would be
useful even if EM counterparts are not identified. Nishizawa et
al. \cite{Nishizawa:2011eq} discussed the possibility of exploiting the
phase shift \cite{Seto:2001qf} of binary sources to constrain
cosmological parameters without EM counterparts. Messenger and 
Read \cite{Messenger:2011gi} proposed to use tidal effects on merging
neutron stars to break the degeneracy to obtain information on the
redshift of the system. Another approach is to take advantage of
the clustering property of GW sources. Recently, Namikawa et
al. \cite{Namikawa:2015prh} showed that the large-angle
auto-correlation of GW sources can provide tight constraints on
primordial non-Gaussianity.

In this paper, we propose the cross-correlation between spatial
distributions of compact binary GW sources without redshift information
and a galaxy sample with known redshifts as a new method to extract
cosmological information from GWs. We show that this
cross-correlation extracts information on the distance-redshift
relation and hence constrains cosmological parameters including the
Hubble constant that determines the absolute distance scale. This is
made possible because the cross-correlation signal is maximized when
luminosity distances of GW sources and redshifts of the galaxy sample
matches. The idea is similar to the cross-correlation of photometric
and spectroscopic galaxies to calibrate photometric redshifts of
galaxies \cite{Newman:2008mb}. A complication is that luminosity
distances estimated from GWs are affected by weak gravitational
lensing, which induces additional spatial correlations on the sky
(e.g., \cite{Camera:2013xfa,Namikawa:2015prh}). In this paper we
explicitly include this effect in our formulation.

This paper is organized as follows. In Section~\ref{sec:formulation}
we give our formulation. We present our result including Fisher matrix
analysis in Section~\ref{sec:result}. Finally we give our conclusion
in Section~\ref{sec:conclusion}. Our fiducial cosmological model is
based on the latest Planck result \cite{Ade:2015xua} and has
matter density $\Omega_{\rm m}=0.308$, dark energy density
$\Omega_{\rm de}=0.692$, baryon density $\Omega_{\rm b}=0.04867$, the
dimensionless Hubble constant $h=0.6763$, the spectral index $0.9677$,
the normalization of matter fluctuations $\sigma_8=0.815$, and dark
energy equation of state $w_{\rm de}=-1$. Throughout the paper we
assume a flat Universe. 

\section{Auto- and cross-correlations of gravitational wave sources}
\label{sec:formulation}

\subsection{Distance to gravitational wave sources}
\label{sec:distance}
Observations of GWs from mergers of compact binaries provide
information on luminosity distances $D$ to the GW sources. On the
other hand, their redshifts are not a direct observable. In this
paper, for simplicity, we assume that the observed luminosity distance
$D_{\rm obs}$ from the analysis of the waveform is related to the true
distance $D$ via the log-normal distribution  
\begin{equation}
p(D_{\rm obs}|D)=\frac{1}{\sqrt{2\pi}\sigma_{\ln D}}
\exp\left[-x^2(D_{\rm obs})\right]\frac{1}{D_{\rm obs}},
\label{eq:dobs}
\end{equation}
where
\begin{equation}
x(D_{\rm obs})\equiv \frac{\ln D_{\rm obs}-\ln D}{\sqrt{2}\sigma_{\ln D}}.
\end{equation}
The dispersion on $D_{\rm obs}$ is caused by various effects,
including the statistical error of GW observations and the degeneracy
of the luminosity distance with other parameters such as masses of
compact objects and the inclination of the system. The parameter
$\sigma_{\ln D}$ quantifies the dispersion. Considering Einstein
Telescope \cite{Punturo:2010zz} like GW observations, in this paper we
assume the dispersion of the distance estimate of $\sigma_{\ln
  D}=0.05$ (e.g., \citep{Camera:2013xfa}). In addition, weak
gravitational lensing also affects cosmological distances. We include
the effect of weak gravitational lensing explicitly by relating the
distance $D$ to the average distance $\bar{D}$, which represents the
standard luminosity distance computed from the homogeneous and
isotropic Friedmann-Robertson-Walker Universe, as  
\begin{equation}
D=\bar{D}\mu^{-1/2}\approx \bar{D}\left[1-\kappa(\bmf{\theta},z)\right],
\label{eq:dis_kappa}
\end{equation}
where $\kappa(\bmf{\theta},z)$ is the lensing convergence, which is a
function of the sky position $\bmf{\theta}$ and redshift $z$. The lensing
convergence is essentially a projected matter density field, and is
given by 
\begin{eqnarray}
\kappa(\bmf{\theta},z)&=&\int_0^z dz'
\frac{\bar{\rho}_{\rm m}(z')}{H(z')(1+z')\Sigma_{\rm crit}(z';z)}\delta_{\rm
  m}(\bmf{\theta},z)\nonumber\\
&\equiv&\int_0^z dz'\,W^\kappa (z';z) \delta_{\rm m}(\bmf{\theta},z'),
\end{eqnarray}
where $\delta_{\rm m}(\bmf{\theta},z)$ is the matter density field, $H(z)$
is the Hubble parameter, $\Sigma_{\rm crit}(z;z_s)$ is the (physical)
critical surface density at redshift $z'$ for the source redshift $z$, and
$\bar{\rho}_{\rm m}(z)=\Omega_{\rm m}\rho_{\rm cr,0}(1+z)^3$ is a mean physical density
of the Universe at redshift $z$.

\subsection{Projected density field of gravitational wave sources}
We construct the $i$-th angular density field of GW sources by
projecting them in the luminosity distance range $D_{{\rm min},i}<D_{\rm obs}<D_{{\rm max},i}$.
Given the log-normal relation between the observed and true distances,
the angular number density is computed from the three-dimensional
number density field of GW sources $n_{\rm GW}(\bmf{\theta},z)$ as 
\begin{eqnarray}
n^{\rm w}_i(\bmf{\theta})=\int_0^\infty
dz\frac{\chi^2}{H(z)} S_{i}(z)\,n_{\rm GW}(\bmf{\theta},z),
\label{eq:numdens}
\end{eqnarray}
where the comoving angular diameter distance is $\chi=\int_0^z
dz'[1/H(z')]$ and $S_{i}(z)$ describes the selection function along
the line-of-sight 
\begin{eqnarray}
S_{i}(z)&\equiv&\frac{1}{2}\left[{\rm erfc}\{x(D_{i,{\rm min}})\}-{\rm
    erfc}\{x(D_{i,{\rm max}})\}\right].
\label{eq:zselect}
\end{eqnarray}
The average projected number density of GW sources in the $i$-th bin
is 
\begin{eqnarray}
\bar{n}^{\rm w}_i&=&\int_0^\infty
dz\frac{\chi^2}{H(z)} S_{i}(z)\,\bar{n}_{\rm GW}(z)\nonumber\\
&=&\int_0^\infty
dz\frac{\chi^2}{H(z)} S_{i}(z)\,T_{\rm obs}\frac{\dot{n}_{\rm GW}(z)}{1+z}.
\label{eq:nave_w}
\end{eqnarray}
Here $T_{\rm obs}$ is the duration of the observation and $\dot{n}_{\rm GW}(z)$ is 
the rate of merger events that can be observed with GW detectors of
interest. The factor $1+z$ in the denominator accounts for the
cosmological time dilation effect.  

The merger rate $\dot{n}_{\rm GW}(z)$ has not yet been constrained
very well. The observation of GW150914 implies the BH-BH merger rate of 
$\dot{n}_{\rm GW}\sim 10^{-6}-10^{-8} h^3{\rm Mpc^{-3}yr^{-1}}$
at the local Universe \cite{Abbott:2016nhf}. Several models predict
that the BH-BH merger rate increases toward higher redshifts (e.g.,
\cite{Dominik:2013tma,Kinugawa:2014zha,Belczynski:2016obo}). We can
also use GWs from neutron star mergers for our cross-correlation
study, and their rate is estimated to be of similar order. Thus in
this paper we assume
$T_{\rm obs}\dot{n}_{\rm GW}=3\times 10^{-6}h^3{\rm Mpc^{-3}}$
over all the redshift range of our interest.

The expression of the projected number density $n^{\rm w}_i(\bmf{\theta})$
allows us to define the projected density field
$\delta^{\rm 2D,w}_i(\bmf{\theta})$, which plays a central role in our
cross-correlation analysis. While the main fluctuation comes from the
three-dimensional distribution of GW sources, $n_{\rm
  GW}(\bmf{\theta},z)=\bar{n}_{\rm GW}(z)\left[1+\delta_{\rm GW}(\bmf{\theta},z)\right]$, 
the convergence in Eq.~(\ref{eq:dis_kappa}) induces additional spatial
fluctuations. Assuming that $\kappa(\bmf{\theta},z)$ is
sufficiently small, we obtain
\begin{eqnarray}
\delta_i^{\rm 2D,w}(\bmf{\theta})&\equiv&\frac{n^{\rm
    w}_i(\bmf{\theta})-\bar{n}^{\rm w}_i}{\bar{n}^{\rm
    w}_i}\nonumber\\
&\approx&\frac{1}{\bar{n}^{\rm w}_i}\int_0^\infty dz\frac{\chi^2}{H(z)} \bar{n}_{\rm GW}(z) S_{i}(z)\,\delta_{\rm
  GW}(\bmf{\theta},z)\nonumber\\
&&+\frac{1}{\bar{n}^{\rm w}_i}\int_0^\infty dz\frac{\chi^2}{H(z)} \bar{n}_{\rm GW}(z)
T_{i}(z)\,\kappa(\bmf{\theta},z),\nonumber\\
&\equiv & \int_0^\infty dz\,\left[W^{\rm s}_i(z)\delta_{\rm GW}(\bmf{\theta},z)+W^{\rm t}_i(z)\kappa(\bmf{\theta},z)\right],
\label{eq:delgw}
\end{eqnarray}
where
\begin{eqnarray}
T_{i}(z)&\equiv&\frac{-\exp\left[-x^2(D_{i,{\rm min}})\right]+\exp\left[-x^2(D_{i,{\rm max}})\right]}{\sqrt{2\pi}\sigma_{\ln D}}.
\label{eq:zselect_der}
\end{eqnarray}
The first term of Eq.~(\ref{eq:delgw}) describes the intrinsic spatial
inhomogeneity of GW sources, whereas the second term of
Eq.~(\ref{eq:delgw}) comes from the apparent modulation of the
distribution of GW sources on the sky due to weak gravitational
lensing which changes luminosity distances inferred from
waveforms. Eq.~(\ref{eq:zselect_der}) implies that the second term of
Eq.~(\ref{eq:delgw}) is smaller than the first term, but as we will
show later, the second term can make dominate contributions to
correlation signals, because the convergence contains accumulated
information along the line-of-sight. 

\subsection{Angular correlation}
For simplicity we assume a linear bias $\delta_{\rm GW}=b_{\rm
  GW}\delta_{\rm m}$ for GW sources. This assumption is reasonable in
the sense that mergers of compact binary objects are expected to 
be associated with galaxies which are known to trace large-scale
structure of the Universe. The angular power spectrum
of the density field $\delta^{\rm 2D,w}_i(\bmf{\theta})$ in
Eq.~(\ref{eq:delgw}) between $i$-th and $j$-th bins is calculated
using the Limber's approximation \cite{Limber:1954zz,LoVerde:2008re} as 
\begin{eqnarray}
C^{{\rm w}_i{\rm w}_j}(\ell)=C^{{\rm s}_i{\rm s}_j}(\ell)+C^{{\rm s}_i{\rm t}_j}(\ell)+C^{{\rm s}_j{\rm t}_i}(\ell)+C^{{\rm t}_i{\rm t}_j}(\ell),
\label{eq:clww}
\end{eqnarray}
\begin{eqnarray}
C^{{\rm s}_i{\rm s}_j}(\ell)=\int_0^\infty dz\,W^{\rm s}_i(z)W^{\rm s}_j(z)\frac{H(z)}{\chi^2}b_{\rm GW}^2P_{\rm m}\left(\frac{\ell+1/2}{\chi};z\right),
\end{eqnarray}
\begin{eqnarray}
C^{{\rm s}_i{\rm t}_j}(\ell)&=&\int_0^\infty  dz\,W^{\rm t}_j(z)
\int_0^{z} dz'\,W^{\rm s}_i(z')W^\kappa(z';z)
\nonumber\\
&& \times\frac{H(z')}{\chi'{}^2}b_{\rm GW}P_{\rm m}\left(\frac{\ell+1/2}{\chi'};z'\right),
\end{eqnarray}
\begin{eqnarray}
C^{{\rm t}_i{\rm t}_j}(\ell)&=&\int_0^\infty  dz\,W^{\rm t}_i(z) \int_0^\infty  dz'\,
W^{\rm t}_j(z') \int_0^{{\rm min}(z,z')} dz''\nonumber\\
&&\hspace*{-10mm}\times W^\kappa(z'';z)W^\kappa(z'';z')\frac{H(z'')}{\chi''{}^2}P_{\rm m}\left(\frac{\ell+1/2}{\chi''};z''\right),
\end{eqnarray}
where $P_{\rm m}(k;z)$ is the matter power spectrum. Since we are
interested in relatively large angular scales ($\ell\lesssim 300$),
the cross spectrum is dominated by the so-called two-halo term (see
e.g., \cite{Oguri:2010vi}), which suggests that we can use the linear
matter power spectrum for $P_{\rm m}(k;z)$ in $C^{{\rm s}_i{\rm s}_j}$
and $C^{{\rm s}_i{\rm t}_j}$.  On the other hand, $C^{{\rm t}_i{\rm
    t}_j}$ is given by a projection of all matter fluctuations along
the line-of-sight which mixes small and large scale fluctuations. Thus it
may be more appropriate to use the nonlinear matter power spectrum for
$P_{\rm m}(k;z)$ in $C^{{\rm t}_i{\rm t}_j}$.  In this paper, we
compute the transfer function of the linear matter power spectrum
using the result in \cite{Eisenstein:1997ik}, and the nonlinear matter
power spectrum using the result in \cite{Takahashi:2012em}.

\subsection{Cross-correlation with spectroscopic galaxies}
Next we consider a spectroscopic galaxy sample in the $i$-th bin defined by
the redshift range $z_{{\rm min},i} <z<z_{{\rm max},i}$
\begin{eqnarray}
\delta_i^{\rm 2D,g}(\bmf{\theta})= \int_0^\infty dz\,
W^{\rm g}_i(z)\delta_{\rm g}(\bmf{\theta},z),
\end{eqnarray}
where
\begin{eqnarray}
 W^g_i(z)\equiv \frac{1}{\bar{n}^{\rm g}_i}\frac{\chi^2}{H(z)} \bar{n}_{\rm g}(z) \Theta(z-z_{{\rm
     min},i})\Theta(z_{{\rm max},i}-z).
\end{eqnarray}
Here the three-dimensional comoving number density of the spectroscopic
galaxy sample is denoted by $\bar{n}_{\rm g}(z)$, and the average
projected number density in the $i$-th bin is simply computed as
\begin{eqnarray}
 \bar{n}^{\rm g}_i=\int_0^\infty dz\,W^g_i(z).
\label{eq:nave_g}
\end{eqnarray}
In this paper we simply assume a constant number density of $\bar{n}_{\rm
  g}=10^{-3}h^3{\rm Mpc^{-3}}$ which resembles e.g., a spectroscopic
galaxy sample obtained by Euclid \cite{Laureijs:2011gra}. Using the Limber's
approximation, the angular power spectrum of spectroscopic galaxies 
between $i$-th and $j$-th bins is given by
\begin{eqnarray}
C^{{\rm g}_i{\rm g}_j}(\ell)=\delta_{ij}\int_0^\infty dz\,\left[W^{\rm g}_i(z)\right]^2\frac{H(z)}{\chi^2}b_{\rm g}^2P_{\rm m}\left(\frac{\ell+1/2}{\chi};z\right),
\label{eq:clgg}
\end{eqnarray}
where we assumed that there is no overlap of redshift ranges between
different redshift bins, and $b_{\rm g}$ is the bias parameter for the
spectroscopic galaxies.

We now consider the cross-correlation between the GW sources and the
spectroscopic galaxies. From Eq.~(\ref{eq:delgw}), we can compute the
cross-correlation power spectrum as
\begin{equation}
C^{{\rm w}_i{\rm g}_j}(\ell)=C^{{\rm s}_i{\rm g}_j}(\ell)+C^{{\rm
    t}_i{\rm g}_j}(\ell),
\label{eq:cross_tot}
\end{equation}
\begin{eqnarray}
C^{{\rm s}_i{\rm g}_j}(\ell)&=&\int_0^\infty dz\,W^{\rm s}_i(z)W^{\rm
  g}_j(z)\frac{H(z)}{\chi^2}\nonumber\\
&&\times b_{\rm GW}b_{\rm g}P_{\rm m}\left(\frac{\ell+1/2}{\chi};z\right),
\label{eq:cross_sg}
\end{eqnarray}
\begin{eqnarray}
C^{{\rm t}_j{\rm g}_j}(\ell)&=&\int_0^\infty  dz\,W^{\rm t}_i(z)
\int_0^z dz' W^{\rm g}_j(z')W^\kappa(z';z)\nonumber\\
&&\times\frac{H(z')}{\chi'{}^2}b_{\rm g}P_{\rm m}\left(\frac{\ell+1/2}{\chi'};z'\right).
\label{eq:cross_tg}
\end{eqnarray}
We use the linear power spectrum for $P_{\rm m}(k;z)$ in both
$C^{{\rm s}_i{\rm g}_j}$ and $C^{{\rm t}_j{\rm g}_j}$. The power
spectrum $C^{{\rm s}_i{\rm g}_j}$ comes from the first term of
  Eq.~(\ref{eq:delgw}) and represents the physical correlation of
  spatial distributions. On the other hand, $C^{{\rm t}_j{\rm g}_j}$,
  which comes from the second term of Eq.~(\ref{eq:delgw}), is the
  correlation of the weak lensing effect on luminosity distances of GW
  sources with spectroscopic galaxies. Since all matter fluctuations
  along the line-of-sight contributes to weak lensing, it induces 
  non-negligible cross-correlations between luminosity and redshift
  bins which are well separated with each other. 

\section{Result}
\label{sec:result}

\subsection{Cross-correlation signal}
\label{sec:crosssig}

First it is useful to study the cross angular power spectrum 
$C^{{\rm w}_i{\rm g}_j}(\ell)$ which is defined in
Eq.~(\ref{eq:cross_tot}). We fix the luminosity distance bin of GW
sources to that corresponds to $0.9<z<1.1$ in our fiducial
cosmological model. On the other hand, we move the central redshift of
the spectroscopic galaxy sample while fixing the bin width 
to $\Delta z=0.1$ in order to see how the cross-correlation signal
changes as a function of the redshift of the spectroscopic galaxy
sample. For bias parameters, we assume a simple parametric form 
$b_{\rm GW}(z)=b_{\rm w1}+b_{\rm w2}/D(z)$ and 
$b_{\rm g}=b_{\rm g1}+b_{\rm g2}/D(z)$, where $D(z)$ is the linear
growth rate, and choose fiducial parameter values as 
$b_{\rm  w1}=b_{\rm w2}=1$ and $b_{\rm  g1}=b_{\rm g2}=1$.

\begin{figure}[t]
\centering
\includegraphics[width=0.4\textwidth]{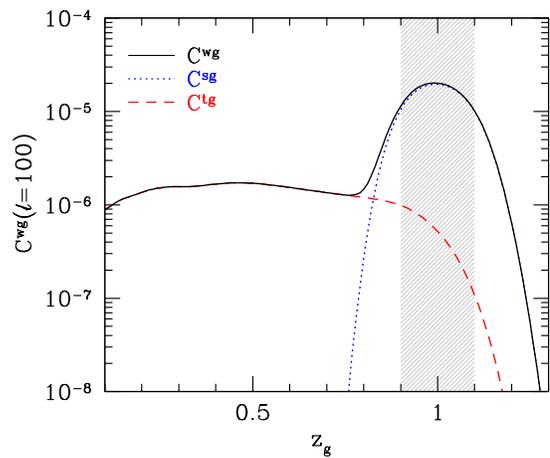}
\caption{The cross-correlation power spectrum $C^{\rm wg}$ between GW
  sources and galaxies [Eq.~(\ref{eq:cross_tot})]. The luminosity
  distance range of GW sources is fixed to that corresponds to
  $0.9<z<1.1$ ({\it gray shaded region}) in our fiducial cosmology. 
  The spectroscopic galaxy sample has the redshift range 
  $z_{\rm g}-\Delta z/2<z< z_{\rm g}+\Delta z/2$ with $\Delta z=0.1$.
  Solid line shows the cross-correlation power spectrum at multipole
  $\ell=100$ as a function of the central redshift of the galaxy
  sample $z_{\rm g}$. Dotted and dashed lines show contributions of
  $C^{\rm sg}$ [Eq.~(\ref{eq:cross_sg})] and $C^{\rm tg}$
  [Eq.~(\ref{eq:cross_tg})] to $C^{\rm wg}$, respectively.
}  
\label{fig:fig1} 
\end{figure}

Fig.~\ref{fig:fig1} shows the cross-correlation power spectrum at
multiple $\ell=100$ as a function of the central redshift of the
spectroscopic galaxy sample $z_{\rm g}$. When the redshift of
the spectroscopic galaxy sample well overlaps with that of GW sources,
the cross-correlation signal becomes large. In this case, the
cross-correlation signal is dominated by the physical correlation of
density fields of GW sources and spectroscopic galaxies, which
corresponds to $C^{\rm sg}$ defined in Eq.~(\ref{eq:cross_sg}).
The cross-correlation signal is maximized when the luminosity distance
bin best matches with the redshift bin, from which we can infer the
relation between the luminosity distance and redshift.
However, Fig.~\ref{fig:fig1} indicates that the cross-correlation
signal extend to much lower redshift of the spectroscopic galaxy
sample. This extra correlation originates from $C^{\rm tg}$ defined in  
Eq.~(\ref{eq:cross_tg}). As stated above, this term represents the
correlation of galaxies and matter fluctuations along the
line-of-sight that induces weak gravitational lensing effect on
luminosity distances of GW sources. We include this large-distance
cross-correlations in our Fisher matrix analysis below.

\subsection{Fisher matrix analysis}

Here we estimate how well we can constrain the distance-redshift
relation and hence cosmological parameters from the cross-correlation
analysis. For this purpose we need the covariance matrix of auto- and
cross-correlation power spectra. Assuming Gaussian statistics, the
covariance matrix is given by 
\begin{eqnarray}
{\rm Cov}\left[C^{ij}(\ell),
  C^{mn}(\ell')\right]&=&\frac{4\pi}{\Omega_{\rm s}}\frac{\delta_{\ell\ell'}}
{(2\ell+1)\Delta\ell}\nonumber\\
&&\times\left(\tilde{C}^{im}\tilde{C}^{jn}+\tilde{C}^{in}\tilde{C}^{jm}\right),
\label{eq:cov}
\end{eqnarray}
where the indices $i$, $j$, $\ldots$ run over $w_i$ and $g_i$,
$\Omega_s$ is the survey area, $\Delta\ell$ is the width of $\ell$
bin, and $\tilde{C}$ denotes the power spectrum including shot noise 
\begin{eqnarray}
\tilde{C}^{ij}=C^{ij}+\delta_{\ij}\frac{1}{\bar{n}_i},
\end{eqnarray}
where $\bar{n}_i$ is the projected number density given by
Eqs.~(\ref{eq:nave_w}) and (\ref{eq:nave_g}). 

With this covariance matrix, we can compute the Fisher matrix as
\begin{eqnarray}
F_{\alpha\beta}=\sum_\ell\sum_{i,j,m,n}\frac{\partial C^{ij}}{\partial
  p_\alpha}\left[{\rm Cov}\left(C^{ij},  C^{mn}\right) \right]^{-1}
 \frac{\partial C^{mn}}{\partial p_\beta},
\label{eq:fisher}
\end{eqnarray}
where $p_\alpha$ denotes cosmological and nuisance parameters.
A marginalized error on each parameter is obtained by
$\sigma(p_\alpha)=\sqrt{\left(F^{-1}\right)_{\alpha\alpha}}$. 

We compute the Fisher matrix with the following setup. For the
correlation among GW sources [$C^{{\rm w}_i{\rm w}_j}$; see
Eq.~(\ref{eq:clww})], while there are weak correlations between
different luminosity distance bins, we ignore them and consider only
correlations between the same luminosity distance bins (i.e., $C^{{\rm
    w}_i{\rm w}_j}\approx\delta_{ij}C^{{\rm  w}_i{\rm w}_i}$). This is
because such weak correlations between different bins are expected not
to affect our results due to relatively large shot noise of GW
sources. The correlation among spectroscopic galaxies
[$C^{{\rm g}_i{\rm g}_j}$; see Eq.~(\ref{eq:clgg})] also does not have
any correlation between different redshift bins. On the other hand, we
consider all the combination of bins for the cross-correlation
[$C^{{\rm w}_i{\rm g}_j}$; see Eq.~(\ref{eq:cross_tot})], given that
there are large-distance correlations as shown in Fig.~\ref{fig:fig1}.

As stated above, observations we have in mind are the Einstein
Telescope \cite{Punturo:2010zz} for GWs and Euclid \cite{Laureijs:2011gra}
for spectroscopic galaxies, although we do not tune our parameters to
these surveys very carefully. We consider the redshift range of
$0.3<z<1.5$, where the maximum redshift mainly comes from the upper
limit of redshifts of the spectroscopic galaxy sample. We define
luminosity distance bins of GW sources by the distance width
corresponding to $\Delta z=0.2$, and compute the minimum
($D_{i,{\rm min}}$) and 
maximum ($D_{i,{\rm max}}$) luminosity distances in each bin using the
standard luminosity distance-redshift relation in our fiducial
cosmological model. Thus we have $N_{\rm w}=6$ luminosity distance bins for
GW sources. On the other hand, we define redshift bins for
spectroscopic galaxies with the interval $\Delta z=0.1$, leading to
$N_{\rm g}=12$ bins. We also consider all the $N_{\rm w}\times N_{\rm
  g}$ cross-correlations for our Fisher matrix analysis. Therefore,
the total number of elements of $C^{ij}$ vector in
Eq.~(\ref{eq:fisher}) is 90. This means that the covariance matrix has
the dimension $90\times90$.

\begin{figure}[t]
\centering
\includegraphics[width=0.35\textwidth]{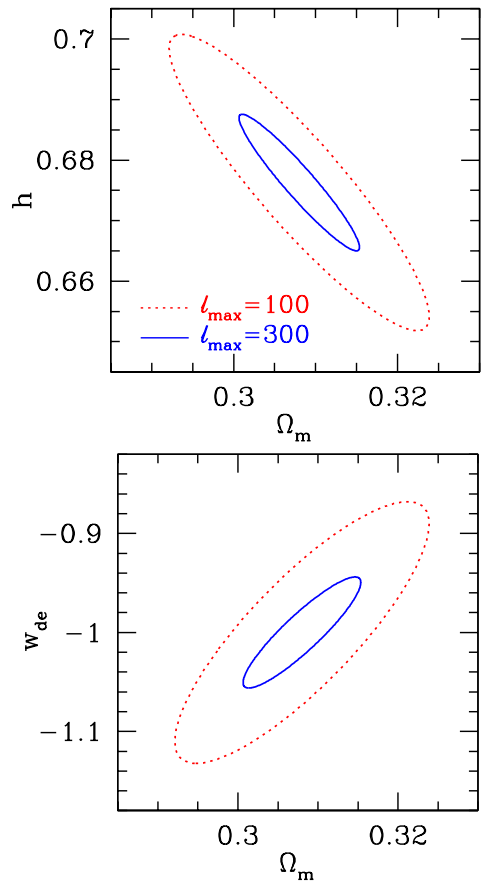}
\caption{
Projected 68\% confidence limit constraints in the $\Omega_{\rm
  m}$-$h$ ({\it top}) and $\Omega_{\rm m}$-$w_{\rm de}$ ({\it bottom})
planes. In each panel, the other model parameters are marginalized
over. Solid lines show constraints for $\ell_{\rm max}=300$, whereas
dotted lines show constraints for $\ell_{\rm max}=100$.
}  
\label{fig:fig2} 
\end{figure}

\begin{table}[t]
\begin{tabular}{cccc}
\hline
\hline
{Model}  &  $\sigma(h)$ &  $\sigma(\Omega_{\rm m})$ & $\sigma(w_{\rm de})$ \\
\hline
$\ell_{\rm max}=100$ & 0.016 & 0.010 & 0.087 \\
$\ell_{\rm max}=300$ & 0.007 & 0.005 & 0.037 \\
\hline
\hline  
\end{tabular}
\caption{Expected marginalized errors on each cosmological
  parameter. See text for details.}
\label{table:par}
\end{table}

As for cosmological parameters, we consider $h$, $\Omega_{\rm m}$, and
$w_{\rm de}$ as parameters controlling the distance-redshift relation.
We also include $\sigma_8$, which determines the normalization of
power spectra, as a parameter. In addition, we include bias parameters 
$b_{\rm  w1}$, $b_{\rm w2}$, $b_{\rm  g1}$, and $b_{\rm g2}$ (see
Sec.~\ref{sec:crosssig}) as nuisance parameters. In total we have 8
parameters that constitute $p_\alpha$ in Eq.~(\ref{eq:fisher}). All
the parameters are treated as free parameters, except $\sigma_8$ for
which we add a weak prior $\sigma(\sigma_8)=0.1$ because we find that
$\sigma_8$ strongly degenerates with the bias parameters.

We consider the multipole range $10<\ell<\ell_{\rm max}$. The maximum
multipole $\ell_{\rm max}$ should be determined from the angular
resolution of GW observations. The angular resolution of the Einstein
Telescope network corresponds to approximately $\ell_{\rm max}=100$
\cite{Namikawa:2015prh,Sidery:2013zua}, which we adopt as a fiducial
value. However we also consider an optimistic case $\ell_{\rm max}=300$
in order to check the dependence of our results on $\ell_{\rm max}$.
Finally we assume the survey area of $\Omega_{\rm s}=15000$~deg$^2$
which is covered by Euclid.

Fig.~\ref{fig:fig2} shows marginalized errors on cosmological
parameters for both $\ell_{\rm max}=100$ and $\ell_{\rm max}=300$. We
find that tight constraints on the Hubble constant as well as dark
energy parameters can be obtained from the cross-correlation
analysis. The summary of constraints given in Table~\ref{table:par}
indicates that a percent level constraint on the Hubble constant can
be obtained, suggesting that the distance-redshift relation is
successfully constrained. We also note a large improvement of
constraints from $\ell_{\rm max}=100$ to $\ell_{\rm max}=300$, which
implies that accurate localizations of GW sources on the sky are
crucial for the cross-correlation analysis.

The expected accuracy of cosmological parameter estimation depends on
several parameters such as the number density of GW sources and their
bias factors, which are poorly known. For comparison, we consider more
pessimistic case with an order of magnitude smaller number density of
GW sources, $T_{\rm obs}\dot{n}_{\rm GW}=3\times 10^{-7}h^3{\rm Mpc^{-3}}$,
and repeat the Fisher matrix calculation. We find that the change of
the expected constraint on the Hubble constant is modest, from
$\sigma(h)=0.016$ to $0.030$ for $\ell_{\rm max}=100$, and from
$\sigma(h)=0.007$ to  $0.013$ for $\ell_{\rm max}=300$. This suggests
that the cross-correlation technique is still useful even when the GW
rate is significantly smaller than our fiducial value.

We note that the expressions of the angular power spectra in this
paper have been derived using the Limber's approximation which breaks
down at small $\ell$ \cite{LoVerde:2008re,Jeong:2009wi}. We expect
that this approximation is valid for the purpose of this paper,
because the cross-correlation signal mainly comes from large $\ell$,
$\ell \sim \ell_{\rm max}$, at which the Limber's approximation is
expected to be reasonably accurate for our choice of $\Delta z=0.1$
for the spectroscopic galaxy sample. Limber's approximation becomes
inaccurate for cross-correlation with large redshift differences, but
due to relative large shot noise such cross-correlation does not
contribute to the result very much. Although there is a long tail of
cross-correlation signals toward lower redshifts
(Fig.~\ref{fig:fig1}), it is essentially the cross-correlation of
galaxies and matter at the same redshift and hence the Limber's
approximation is again accurate. Nevertheless, we caution that the
full calculation without the Limber's approximation may be required for
more accurate predictions of the cross-correlation signals, which is
beyond the scope of this paper.

\section{Conclusion}
\label{sec:conclusion}

GWs from mergers of compact objects such as BHs serve as a useful
cosmological probe because they allow us to directly measure absolute
distance scales. However, in order to constrain the distance-redshift
relation from GW sources we also need redshift information. While the
redshift information may be obtained from observations of EM
counterparts,  it is unclear whether such EM counterparts can 
be reliably identified, especially for BH-BH mergers. In this paper,
we propose to use the cross-correlation of GW sources with
spectroscopic galaxies as an alternative means of constraining the
distance-redshift relation. We have explicitly included the effect of
weak gravitational lensing on luminosity distance estimates in our
formulation. Using the Fisher matrix formalism, we have shown that
tight constraints on the Hubble constant as well as dark energy
parameters can be obtained by the cross-correlation of GW sources
observed by the Einstein Telescope and spectroscopic galaxies observed
by Euclid. 

Constraints on absolute distance scales at cosmological distances
are not directly obtained except a few cases (e.g.,
\cite{Oguri:2006qp,Suyu:2012aa,Eisenstein:2005su,Anderson:2013zyy}). GW
standard sirens therefore offer invaluable information on the
distance-redshift relation including the absolute distance scale. Our
analysis has shown that it is possible to constrain the
distance-redshift relation even without identifying EM counterparts of
GW sources and thus without any redshift information on individual GW
sources.

Finally we note that there is room for improving constraints on the
distance-redshift relation from the cross-correlation analysis.
For instance, while we have restricted our analysis to $z<1.5$, GWs
can be detected out to much higher redshifts in next-generation GW
observations. For those high-redshift GWs, we can use high-redshift
tracers such as quasars, or we may also be able to galaxies with
photometric redshifts for the cross-correlation analysis, if their
photometric redshifts are accurate enough. For example, Euclid's
near-infrared photometry, when combined with ground-based optical
photometry, enables an accurate determination of photometric redshifts
easily out to $z\sim 3$ \cite{Abdalla:2007uc}. Photometric galaxies are
much denser and therefore lead to more significant detections of
cross-correlation signals, which suggests that cross-correlation of GW
sources with galaxies with photometric redshifts has a great potential
to enhance the use of the cross-correlation method proposed in this
paper. The distance-redshift relation may also be
constrained by cross-correlating GW sources with tomographic weak
lensing \cite{Hu:1999ek}. We leave exploring these possibilities for
future work.  

\acknowledgments 
I thank the anonymous referee for useful suggestions. 
This work was supported in part by World Premier International
Research Center Initiative (WPI Initiative), MEXT, Japan, and JSPS
KAKENHI Grant Number 26800093 and 15H05892.

\end{document}